\documentclass[journal]{IEEEtran}

\usepackage{cite}
\usepackage{amsmath}
\usepackage{amssymb}
\usepackage{amsfonts}
\usepackage{graphicx}
\usepackage{textcomp}
\usepackage{xcolor}
\usepackage{array}
\usepackage{subfigure}
\usepackage[font=normalsize]{caption}
\usepackage{stfloats}
\usepackage{url}
\usepackage{verbatim}
\usepackage[export]{adjustbox}
\usepackage{ntheorem}
\usepackage{longtable}
\usepackage{booktabs}
\usepackage{makecell}
\usepackage{mwe}
\usepackage{graphbox}
\usepackage{enumitem}
\usepackage{ragged2e}
\usepackage{multicol} 
\usepackage[colorlinks, linkcolor=black, anchorcolor=black, citecolor=blue]{hyperref}
\usepackage{comment}
\usepackage{color}
\usepackage{bbding} 
\usepackage{diagbox} 
\usepackage{float}
\usepackage{multirow}
\theorembodyfont{\upshape}

\makeatletter
\renewtheoremstyle{plain}
{\item{\theorem@headerfont ##1\ ##2\theorem@separator}~}
{\item{\theorem@headerfont ##1\ ##2\ (##3)\theorem@separator}~}
\makeatother
\def\BibTeX{{\rm B\kern-.05em{\sc i\kern-.025em b}\kern-.08em
		T\kern-.1667em\lower.7ex\hbox{E}\kern-.125emX}}
\usepackage{balance}
\graphicspath{{QUDM/EPS/}}
\usepackage{lipsum}


\usepackage{setspace}
\graphicspath{{}}
\setenumerate[1]{itemsep=0pt,partopsep=0pt,parsep=\parskip,topsep=0pt}
\setitemize[1]{itemsep=0pt,partopsep=0pt,parsep=\parskip,topsep=0pt}
\setdescription{itemsep=0pt,partopsep=0pt,parsep=\parskip,topsep=0pt}

\usepackage{algorithm} 
\usepackage{algorithmicx}  
\usepackage{algpseudocode}  
\begin{document}
\title{HQCC: A Hybrid Quantum-Classical Classifier with Adaptive Structure}	
\author{Ren-Xin Zhao, Xinze Tong, Shi Wang$^*$
		
\IEEEcompsocitemizethanks{
\IEEEcompsocthanksitem{
}
\IEEEcompsocthanksitem{
(corresponding author: Shi Wang)

Ren-Xin Zhao is with the School of Computer Science and Engineering, Central South Univerisity, China, Changsha, 410083 and visiting the School of Computer Science and Statistics, Trinity College Dublin, Ireland, Dublin, D02PN40 (Email: renxin\_zhao@alu.hdu.edu.cn). 

Shi Wang and Xinze Tong are with the College of Electrical and Information Engineering, Hunan University, China, Changsha, 410082 (Email: Shi\_Wang@hnu.edu.cn, 1569125042@qq.com).

IEEE Letters Demo
}
}
}

\maketitle
	
\begin{abstract}
Parameterized Quantum Circuits (PQCs) with fixed structures severely degrade the performance of Quantum Machine Learning (QML). To address this, a Hybrid Quantum-Classical Classifier (HQCC) is proposed. It opens a practical way to advance QML in the Noisy Intermediate-Scale Quantum (NISQ) era by adaptively optimizing the PQC through a Long Short-Term Memory (LSTM) driven dynamic circuit generator, utilizing a local quantum filter for scalable feature extraction, and exploiting architectural plasticity to balance the entanglement depth and noise robustness. We realize the HQCC on the TensorCircuit platform and run simulations on the MNIST and Fashion MNIST datasets, achieving up to 97.12\% accuracy on MNIST and outperforming several alternative methods.
\end{abstract}
	

\section{Introduction}\label{sec1}
\IEEEPARstart{A}dopting PQCs with fixed structures somehow severely restricts the performance of a large portion of QML models \cite{0.0,0.1,0.2,0.3,0.4,0.17,0.18}. Specifically, only adjusting the parameters in fixed-structure PQCs is insufficient to fully exploit their expressive power and entanglement performance, and cannot dynamically align with task-specific feature hierarchies \cite{0.5,0.6,0.6.0,0.7}. This bottleneck is more acute in high-dimensional classification tasks that require multi-scale correlations \cite{0.8}. In addition, the inflexibility of the architecture not only weakens the quantum advantage \cite{0.9} in practical applications, but also destroys the resilience to noise \cite{0.10}, becoming a key challenge for the efficient deployment of QML models in the era of NISQ \cite{0.11}. Fortunately, the Quantum Architecture Search (QAS) can alleviate the above dilemma.

The QAS refers to a method for dynamically optimizing PQC architectures for specific tasks, which is generally divided into gradient-based meta-learning \cite{0.12}, Reinforcement Learning (RL)-driven gate selection \cite{0.13}, and genetic algorithm-based topology optimization \cite{0.14}. Gradient-based methods iteratively prune redundant gates via sensitivity analysis, but are computationally expensive and fail to capture sequential dependencies in quantum feature maps \cite{0.15}. RL-based methods automatically design gate sequences but lack stability under NISQ noise due to sparse reward mechanisms \cite{0.16}. Genetic algorithms explore diverse architectures via mutation and crossover operations, but their heuristic nature prevents convergence guarantees \cite{0.19}. Most importantly, existing AQAS frameworks ignore classical memory mechanisms for stable architecture synthesis, limiting their ability to balance entanglement depth and noise robustness—a key trade-off for NISQ compatibility \cite{0.20}.

This letter proposes a HQCC that addresses these limitations through three innovations: (1) A dynamic circuit generator employing LSTM-controlled \cite{0.21} quantum gate sequences autonomously explores optimal topologies, enhancing entanglement and nonlinear feature extraction. (2) Local quantum filters with sliding-window processing enable shallow PQCs to handle high-dimensional data while maintaining NISQ compatibility. (3) Architectural plasticity allows task-specific adaptation of entanglement depth and gate connectivity, outperforming fixed-topology benchmarks in classification accuracy. By decoupling circuit optimization from parameter training, HQCC establishes a new paradigm for balancing quantum expressivity with NISQ constraints, advancing practical QML deployment.

\section{Methodology}\label{sec3}
This section introduces the proposed HQCC model, which comprises several components. Firstly, we present an overview of HQCC, providing explanations for each component. Subsequently, we delve into a detailed description of the principles and architecture of the quantum node VQCL within HQCC.

\subsection{HQCC}
The HQCC model proposed in this paper adopts a simple structure: $Conv1-Pool1-Conv2-Pool2-VQCL-FC1-FC2$. In this structure. Conv1 and Conv2 are the classical convolutional layers, and Pool1 and Pool2 are the classical pooling layers. The former is used for extracting shallow features from the input data, while the latter is used for compressing and downsizing the features extracted by the former. Then, further feature extraction of the extracted classical features is performed by VQCL. Compared to classical convolutional layer, VQCL utilizes PQC as its kernel function, leveraging the powerful expressive capability and high-dimensional feature extraction capacity of quantum circuits. It is expected to better capture complex relationships and patterns within the data. Finally, the features extracted by VQCL are fed into the fully connected layer. The fully connected layer combines these features with weights, applies an activation function, and generates the ultimate prediction results. If HQCC is denoted as function $f(x;w_c,\theta)$, where $x$ is the input, $w_c$ is the parameter of the classical node of the model, and $\theta$ is the parameter of the quantum node. then the supervised classification learning task can be modeled as:
\begin{equation}
\setlength{\abovedisplayskip}{3pt}\setlength{\belowdisplayskip}{3pt}
\underset{\theta ,{{w}_{c}}}{\mathop{\min }}\,J(\theta ,{{w}_{c}}):=\mathcal{L}[f(x;\theta ,{{w}_{c}})-y],
\end{equation}
where $\mathcal{L}$ is the cross-entropy loss function, which is used to compare the difference between the output of the function $f$ and the true label $y$.
\begin{figure*}[h]
     \centering
         \includegraphics[scale=1]{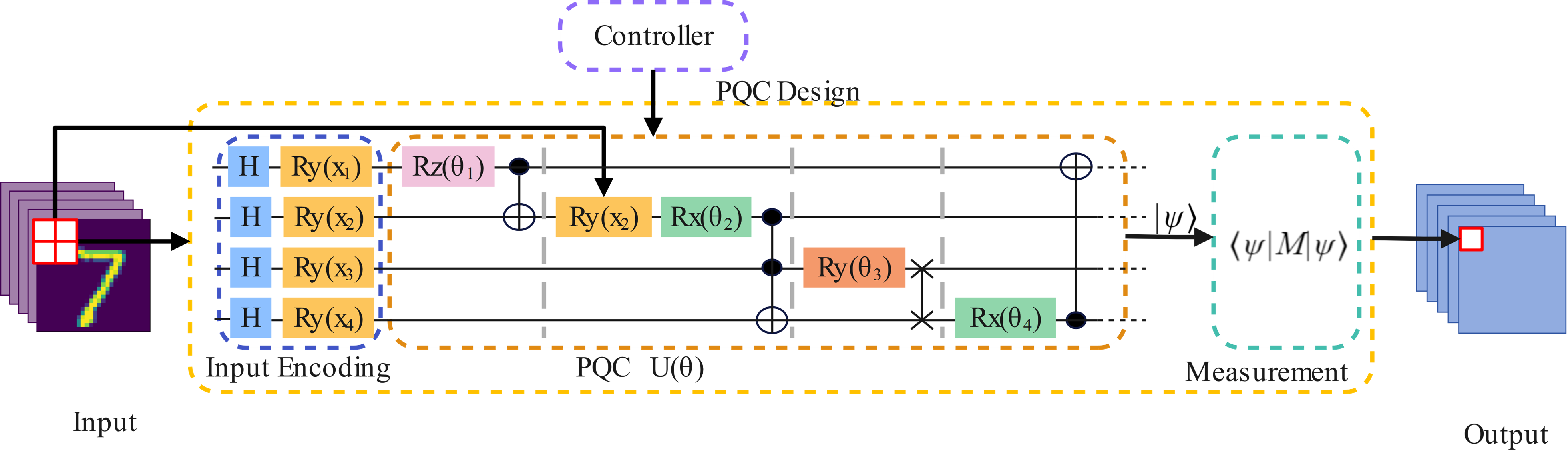}
    \caption{Framework of the QUDM}
    \label{QCNNEDM}
\end{figure*}
\subsection{VQCL}
\subsubsection{Input Encoding}
As mentioned in Section II, Part B, although the amplitude encoding provides higher spatial efficiency, angle encoding is relatively more flexible because it allows adjusting the mapping approach by selecting a suitable mapping function. In addition, the number of qubits required for VQCL depends on the window size, and the required space complexity is $O(mn)<O(N)$, which can reduce the impact of the low space efficiency of angle encoding to a certain extent. Therefore, in this paper, the conversion from classical data to quantum data is achieved by angle encoding. The classical data is normalized to between 0 and 1 before being input to VQCL. Fig. 1 shows encoding four classical data in a $2\times2$ sliding window into four qubits using single-qubit rotation gates $R_y$. Before data encoding, we apply Hadamard gates $H$ to all qubits in the base state $|0\rangle$ to produce a balance superposition of all the base states. The whole process can be represented as:
\begin{equation}
\setlength{\abovedisplayskip}{3pt}\setlength{\belowdisplayskip}{3pt}
x\to |\varphi (x)\rangle =\underset{i=1}{\overset{N}{\mathop{\otimes }}}\,{{R}_{y}}({{a}_{i}})H|0\rangle.
\end{equation}
\subsubsection{Circuit Architecture}
VQCL utilizes a PQC as a filter for pattern learning and feature extraction from the data. The design of PQC adopts a multi-layer architecture, and only one layer is shown in Fig. 1. Assume that the input data to the quantum filter is a normalized two-dimensional array X of size $Q\times P$, and a sliding window of size m×n is defined with a step size of s, where $m\leq Q$ and $n\leq P$. The quantum filters directly encode classical data x within a sliding window into input quantum states $|\varphi (x)\rangle $ using angle encoding. $|\varphi (x)\rangle $ is then evolved in the PQC to obtain the output quantum state $|\varphi (x)\rangle_{out} $. The process can be expressed as:
\begin{equation}
\setlength{\abovedisplayskip}{3pt}\setlength{\belowdisplayskip}{3pt}
|\varphi (x){{\rangle }_{out}}=U(\theta )|\varphi (x)\rangle ,
\end{equation}
where
\begin{equation}
\setlength{\abovedisplayskip}{3pt}\setlength{\belowdisplayskip}{3pt}
U(\theta )=\prod\limits_{l=1}^{L}{\underset{i=1}{\overset{N}{\mathop{\otimes }}}\,\prod\limits_{j=1}^{J_{i}^{l}}{U_{i,j}^{l}(\theta _{i,j}^{l})}}.
\end{equation}
$N$ is the dimension of the classical data vector $x$ within the sliding window, $L$ is the number of layers of the PCQ, $J_i^l$ is the number of parameterized unitaries on the $i$-th qubit of the $l$-th layer of the PQC, and $\theta _{i,j}^{l}$ is the parameter of the parameterized unitaries at the corresponding position. To make it clear here: the non-parameterized unitaries are regarded as the parameterized unitaries whose parameter $\theta$ is constant and uniformly denoted by $U_{i,j}^{l}(\theta _{i,j}^{l})$ for the convenience of representation. In the end, the Pauli-Z measurement operator $\delta_z^r$ is utilized to measure the readout qubit r. For a multiqubit system, the measurement operator M can be constructed using 
\begin{equation}
\setlength{\abovedisplayskip}{3pt}\setlength{\belowdisplayskip}{3pt}
M=\delta _{0}^{1}\otimes \cdots \otimes \delta _{z}^{r}\otimes \cdots \otimes \delta _{0}^{N},
\end{equation}
where the superscript denotes the index of the qubits. Apart from the readout qubit r, no measurements are conducted on other qubits, and the measurement operators on these qubits are represented by the identity matrix $\delta_0^i$. The measurement outcome can be obtained using 
\begin{equation}
\setlength{\abovedisplayskip}{3pt}\setlength{\belowdisplayskip}{3pt}
E(\theta )={{\langle \varphi (x){{|}_{out}}M|\varphi (x)\rangle }_{out}}.
\end{equation}

Researches show that the architecture of quantum circuits affects their expressive and entanglement capabilities [27]. As for this problem, this paper draws on inspiration from the work of et al [25]. and uses a controller to generate PQC architecture designs suitable for HQCC. As shown in Fig. 2. The controller is actually a deep neural network (DNN) without inputs. Based on the work of Mohammad Pirhooshyaran et al. [25], this paper adds LSTM layer to the DNN[35]. LSTM is a special kind of recursion neural network (RNN) which solves the problem of gradient vanishing and gradient exploding in the standard RNN [36], and has better long-term memory capability. As an internal memory unit in the controller, LSTM helps the controller learn and capture the nonlinear relationships between quantum gates during the generation process of PQC architecture design. By memorizing the state of a previously generated PQC architecture, LSTM is able to guide the generation of the next PQC architecture design based on this state. Before starting the training of the HQCC, according to the theory of deep reinforcement learning, the controller needs to output a policy $\pi_t(w_d)$ for all decisions to be made for each qubit in each layer of the PQC, which is parameterized by the random weights $w_d$ of the DNN. Then, samples drawn from $\pi_t(w_d)$  are taken as the proposed PCQ architecture design.
\begin{figure}[h]
     \centering
         \includegraphics[scale=0.8]{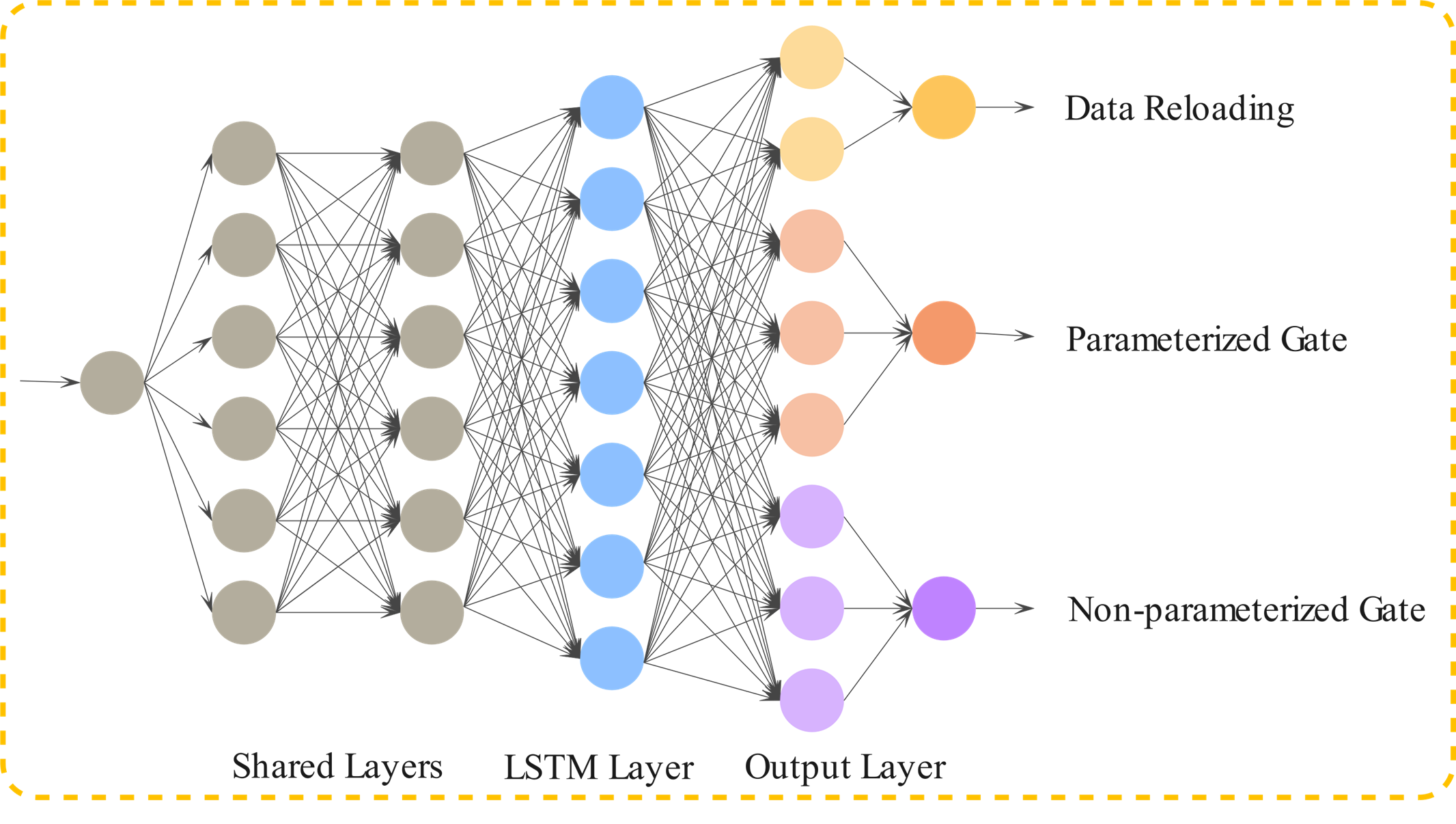}
    \caption{Schematic diagram of controller.}
    \label{QCNNEDM1}
\end{figure}

Fig. 2 shows that decisions are categorized into three types: data reloading, parameterized operations, and non-parameterized operations. The decision to reload data mitigates, to some extent, the exponential vanishing of the cost function with respect to the gradient of the circuit parameters [37]. In this paper, parameterized operations include {Rx, Ry, Rz, CRx, CRy, CRz}. Non-parameterized operations include {H, X, CNOT, SWAP, Toffoli}. Fig. 3 illustrates the training process of HQCC for a supervised classification task. The PQC is architected using the design recommended by the controller. After that, HQCC is trained on the training set. Drawing upon the methods of DRL, when HQCC is trained for a period of time t and the performance stabilizes, the performance metrics (e.g., loss or accuracy) on the test data set are fed back to the controller as signals. The controller then combines the feedback signal and the policy $\pi_t (w)$ characteristics (e.g., entropy) to update its own parameters $w_d$ , and optimize the policy $\pi_t (w)$ . By sampling the new strategy, we can obtain a new design suggested by the controller that improves on the previous PQC architecture design. HQCC constructs the PQC according to the new design and retrains. The process is repeated until the performance of HQCC meets our requirements. In this way, the controller can adaptively design the PQC architecture for HQCC for specific tasks.
\begin{figure}[h]
     \centering
         \includegraphics[scale=0.8]{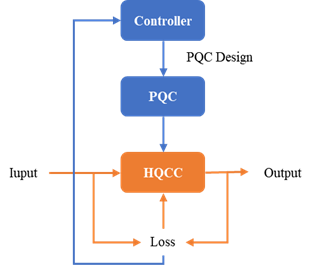}
    \caption{Schematic diagram of controller.}
    \label{QCNNEDM2}
\end{figure}

\section{Experiments and Data Analysis}\label{sec4}
In this section, the classification performance of HQCC (including HQCC$^a$ and HQCC$^b$), VCNN \cite{0.17}, QuanvNN \cite{0.18}, and CNN \cite{3.0} on the MNIST \cite{3.1} and Fashion MNIST \cite{3.2} datasets is evaluated and compared on TensorCircuit \cite{3.3}, an open source platform that supports automatic differentiation, multi-framework integration (TensorFlow/PyTorch/JAX), and GPU/TPU acceleration.
MNIST and Fashion MNIST are broadly adopted benchmark datasets in machine learning, serving as cornerstones for numerous studies in the field. They both consist of 10000 test images and 60000 training images, each containing $28\times28$ pixel points.

\subsection{Binary Classification}
According to Tab. \ref{tab1}, this experiment selects 1000 images from the MNIST dataset as the training set and another 200 images as the test set, and constructs 5 groups of binary classification tasks (0 \& 5, 1 \& 4, 2 \& 8, 3 \& 7, 6 \& 9). All models use a sliding filter of size $3\times 3$ (9 qubits are required for quantum implementation), trained for 30 rounds with a batch size of 64, and the optimizer is Adam with a learning rate of 0.005. The difference is that HQCC uses a single quantum filter (4 for VCNN and QuanvNN). Since the quantum node input feature map data volume is large, reducing the number of filters can reduce the simulation calculation burden. HQCC needs to optimize the PQC architecture through the controller. The controller uses a single-layer LSTM module with 64 hidden units, the time step is synchronized with the filter sliding window. The dropout rate is 0.2. The optimizer is Adam. The learning rate is 0.001.

According to Tab. \ref{tab1} and Tab. \ref{tab2}, the HQCC has significant advantages in many indicators: (1) HQCC$^b$ performed outstandingly. Among the 5 groups of tasks, it achieved the highest accuracy (99.68\%, 99.86\%, 98.34\%) and F1 score (99.88\%, 98.89\%, 99.12\%) in 0 \& 5, 1 \& 4, and 2 \& 8, respectively. Its AUC values (99.90\%, 99.60\%, 99.86\%) are also close to or exceeded that of other models, indicating that the adaptive PQC architecture can dynamically optimize the classification boundary. (2) Compared with VCNN, HQCC$^a$ uses only 45 parameters (VCNN has 108), while achieving a higher average accuracy (99.26\% vs. 98.29\% for 0 \& 5) and a lower standard deviation (such as $\pm 0.13$ vs. $\pm0.10$ for 2 \& 8), proving that the dynamic architecture improves model stability while reducing quantum resources. (3) Compared to CNN, although CNN leads in accuracy/F1 score (98.37\%/98.42\%) in 6 \& 9, 
its AUC (96.47\%) for 0\& 5 and F1 score (98.69\%) for 
1 \& 4 are lower than the HQCC$^a$.
The standard deviation fluctuates significantly (such as $\pm 0.44$ for 2 \& 8), highlighting the advantages of extraction of quantum features. (4) By introducing the LSTM controller, HQCC$^b$ improves the accuracy by 0.42\% (99.68\% vs. 99.26\%) and the F1 score by 0.24\% (99.88\% vs. 99.64\%) in 0/5 compared with HQCC$^a$, indicating that the memory module can enhance the expressiveness of the PQC architecture design and further optimize the performance. In summary, HQCC demonstrates high accuracy and strong robustness with low parameter count through an adaptive PQC architecture, and its design provides an efficient and scalable solution for quantum machine learning tasks.

\begin{table*}[h]
		\centering
		\def\tablename{Tab.}
		\caption{Comparison of test accuracies of the binary classifications}
		\label{tab1}
		\begin{tabular}{@{}lcccccc@{}}
			\toprule
			\multirow{2}{*}{models} & \multirow{2}{*}{parameters} & \multicolumn{5}{c}{label}                      \\ \cmidrule(l){3-7} 
			&                            & 0 \& 5     & 1 \& 4 & 2 \& 8 & 3 \& 7 & 6 \& 9 \\ \midrule
			VCNN      \cite{0.17}              & 108                        & 98.29$\pm$0.18 &    97.17$\pm$0.26    &   97.61$\pm$0.10     &    \textbf{99.36}$\pm$0.18    &   95.70$\pm$0.18     \\
			QuanvNN  \cite{0.18}               & -                          &     96.05$\pm$0.33       &  95.46$\pm$0.22     & 97.51$\pm$0.13        &     96.66$\pm$0.11   &   92.34$\pm$0.25      \\
			CNN  \cite{3.0}                   & -                          &   96.53$\pm$0.33         &    99.72$\pm$0.28     &   97.55$\pm$0.44      &     97.89$\pm$0.19    &   \textbf{98.37}$\pm$0.21     \\
			HQCC$^a$                    & 45                         &       99.26$\pm$0.22     &  99.68$\pm$0.26       & 98.20$\pm$0.13        &    99.17$\pm$0.23     &   98.17$\pm$0.17     \\
			HQCC$^b$                    & 45                         &   \textbf{99.68}$\pm$0.24         &  \textbf{99.86}$\pm$0.15      &  \textbf{98.34}$\pm$0.11       &     99.21$\pm$0.19   &   98.27$\pm$0.07      \\ \bottomrule
            \multicolumn{4}{l}{\footnotesize * HQCC$^a$ does not add LSTM to the controller, while HQCC$^b$ does.} \\
		\end{tabular}
	\end{table*}
	
	\begin{table}[h]
		\centering
		\def\tablename{Tab.}
		\caption{Comparison of test accuracies of the binary classifications}
		\label{tab2}
        \resizebox{\columnwidth}{!}{
		\begin{tabular}{@{}llccccc@{}}
			\toprule
			\multirow{2}{*}{models} & \multirow{2}{*}{models} & \multicolumn{5}{c}{label}\\ \cmidrule(l){3-7}& & 0 \& 5& 1 \& 4& 2 \& 8& 3 \& 7& 6 \& 9\\ \midrule
			\multirow{5}{*}{AUC}   
			&VCNN&99.91&99.62&99.70&\textbf{99.98}&99.41\\
			&QuanvNN&99.73&99.63&99.82&99.35&\textbf{99.88}\\
			&CNN&96.47&99.95&99.79&98.41&99.84\\
			&ASQCENN$^a$&99.84&99.92&99.86&99.87&99.87\\
			&ASQCENN$^b$&99.90&99.60&99.86&99.97&99.84\\ \midrule
			\multirow{5}{*}{F1-score} & VCNN&98.23&97.09&97.56&\textbf{99.37}&95.86\\
			& QuanvNN&95.70&95.05&97.48&96.67&92.90\\
			& CNN&96.23&98.69&97.50&97.87&\textbf{98.42}\\ 
			& HQCC$^a$&99.64&98.77&98.62&99.32&98.16\\ 
			& HQCC$^b$&\textbf{99.88}&\textbf{98.89}&\textbf{99.12}&99.21&98.37\\  \bottomrule
		\end{tabular}
        }
	\end{table}
\subsection{Multi-Class Classification}
This experiment takes 1500 images and an additional 300 images from the MNIST and Fashion MNIST datasets as training and test sets, respectively. For other parameter configurations, please refer to the binary classification experiment.

According to Tab. \ref{tab3}, the following conclusions can be drawn. (1) High-precision robustness: The accuracy on the MNIST dataset is 97.12\% (standard deviation $\pm$0.23), and the recall and F1-score are both 0.9696. The three indicators are highly consistent, indicating that the model has strong stability for digital feature extraction;
(2) Adaptability to complex data: In Fashion MNIST, which has more complex textures, HQCC still maintains an accuracy of 94.10\% ($\pm$0.34) and a recall of 0.9397, proving that it can handle diverse patterns, but it is about 3\% lower than the MNIST indicators (such as accuracy 97.12\% $\to$ 94.10\%), reflecting the challenge of increasing differences within the clothing category;
(3) Error sensitivity: The standard deviation of Fashion MNIST ($\pm$0.34) is higher than that of MNIST ($\pm$0.23), indicating that the abstract features of the clothing category may cause prediction fluctuations, and feature encoding needs to be further optimized;
(4) Comprehensive performance balance: The F1-score of the two datasets (MNIST 0.9696, Fashion MNIST 0.9385) is almost the same as the recall rate, indicating that HQCC achieves an effective trade-off between precision and coverage without significant bias.
(5) Limitations: The confusion matrix shows that the misclassification rate of easily confused categories in Fashion MNIST is high, suggesting that the current quantum feature extraction has limited ability to distinguish fine-grained textures. In the future, it is necessary to explore dynamic filters or hybrid classical-quantum hierarchical structures to improve the ability to capture complex patterns.
\begin{table}[h]
		\centering
		\def\tablename{Tab.}
		\caption{Comparison of test accuracies of the multi-classifications}
		\label{tab3}
		\begin{tabular}{@{}lccc@{}}
			\toprule
			Datasets & Accuracy (\%) & Recall  & F1-score \\ \midrule
			MNIST&  97.12$\pm$0.23&0.9696  & 0.9696 \\
			Fashion MNIST&94.10$\pm$0.34&0.9397&0.9385 \\ \bottomrule
		\end{tabular}
	\end{table}
\section{Conclusions}\label{sec5}
We improve the approach in reference [25] and propose a HQCC algorithm, HQCC. The HQCC algorithm combines the advantages of quantum computing and classical machine learning algorithms, and the experiment shows that it performs well on supervised classification tasks. Since the HQCC algorithm employs the idea of sliding filters, which operate only on localized subregions of the input data, quantum machine learning can be implemented on large datasets with fewer qubits. In addition, HQCC can adaptively design PQC with higher expressive architecture for different quantum machine learning tasks, which can allow us to solve real-world problems better with quantum computers in the NISQ era.

The HQCC demonstrates significant advantages in QML through its adaptive PQC architecture. By dynamically optimizing PQC structures for specific tasks, HQCC achieves high classification accuracy and robustness with reduced quantum resource demands. The integration of LSTM controllers further enhances performance, enabling efficient feature representation and boundary optimization. Notably, the HQCC maintains competitive metrics in complex multi-class scenarios, showcasing scalability across datasets with varying complexity. While challenges remain in fine-grained texture discrimination, the HQCC establishes a foundation for adaptable and resource-efficient quantum machine learning solutions.

\end{document}